\documentclass[11pt]{article}
\usepackage[numbers]{natbib}
\usepackage{rotating}
\usepackage{floatpag}
\rotfloatpagestyle{empty}
\usepackage{amsmath}

\usepackage{amsthm}
\usepackage{graphicx}

\begin{document}

\author{David W. Snoke\\
Department of Physics and Astronomy, University of Pittsburgh\\ 
Pittsburgh, Pennsylvania 15260, USA\\
\\
Andrew J. Daley\\
Department of Physics and Scottish Universities Physics Alliance\\ 
University of Strathclyde, Glasgow G4 0NG, Scotland, UK}

\title{The Question of Spontaneous Symmetry Breaking in Condensates}

\date{}

\maketitle
\begin{abstract}
The question of whether Bose-Einstein condensation involves spontaneous symmetry breaking is surprisingly controversial. We review the theory of spontaneous symmetry breaking in ferromagnets, compare it to the theory of symmetry breaking in condensates, and discuss the different viewpoints on the correspondence to experiments. These viewpoints include alternative perspectives in which we can treat condensates with fixed particle numbers, and where coherence arises from measurements. This question relates to whether condensates of quasiparticles such as polaritons can be viewed as ``real'' condensates.
\end{abstract}

\section{Introduction}
Spontaneous symmetry breaking is a deep subject in physics with long historical roots. At the most basic level, it arises in the field of cosmology. Physicists have long had an aesthetic principle that leads us to expect symmetry in the all of the basic equations of physical law. Yet the universe is manifestly full of asymmetries. How does a symmetric system acquire asymmetry merely by evolving in time? Starting in the 1950's, cosmologists began to borrow the ideas of spontaneous symmetry breaking from condensed matter physics, which were originally developed to explain spontaneous magnetization in ferromagnetic systems. 

Spontaneous coherence in all its forms (e.g. Bose-Einstein condensation, superconductivity, and lasing) can be viewed as another type of symmetry breaking. The Hamiltonian of the system is symmetric, yet under some conditions, the energy of the system can be reduced by putting the system into a state with asymmetry, namely, a state with a common phase for a macroscopic number of particles. The symmetry of the system implies that it does not matter what the exact choice of that phase is, as long as it is the same for all the particles. 

It is not obvious, however, whether the symmetry breaking which occurs in spontaneous coherence of the type seen in lasers or in Bose-Einstein condensation is the same as that seen in ferromagnetic systems. There are similarities in the systems which encourage the same view of all types of symmetry breaking, but there are also differences. In fact, there is a substantial school of thought that symmetry breaking in Bose-Einstein Condensates with ultracold atoms is a ``convenient fiction'' (a term applied to optical coherence by M{\o}lmer \cite{Molmer1997}). That is, in contrast to ferromagnets, we do not have direct experimental access to observe symmetry breaking itself, and the experimental consequences of this theory can be equally reproduced in theories using fixed atom number  \cite{Castin1997,Gardiner1997,Castin1998,Leggett2001,Leggett1991,Javanainen1996,Naraschewski1996,
Wong1996,Wright1996,Stenholm2002,Schachenmayer2011}, without spontaneous symmetry breaking.

In what follows we will review the theory of spontaneous symmetry breaking as applied to ferromagnets, and then discuss the different viewpoints on spontaneous symmetry breaking in condensates. We will touch on theories involving fixed particle numbers, where coherence arises in the measurement process, and also discuss the relationship to the question of whether polariton systems can be ``real'' condensates.

\section{Review of Elementary Spontaneous Symmetry Breaking Theory}
\label{sect.elem}

The canonical example in condensed matter physics for spontaneous symmetry breaking is the ferromagnetic spin system, represented in simple form by the Ising Hamiltonian for a lattice of localized electrons,
\begin{eqnarray}
H = \alpha B \sum_i \sigma_i  - J \sum_{\langle i,j\rangle} \sigma_i\sigma_j ,
\label{ising}
\end{eqnarray}
where $\sigma_i = a^{\dagger}_{i\uparrow} a_{i\uparrow}-a^{\dagger}_{i\downarrow} a_{i\downarrow}$ is the spin operator for site $i$ and the sum $\langle i,j\rangle$ is for nearest neighbors. The first term gives the effect of an external magnetic field $B$, and the second term the effect of spin interactions which favor alignment. The order parameter for the system is defined as
\begin{equation}
m = \frac{1}{N} \sum_i \sigma_i,
\end{equation}
which is the average magnetization. More generally, in a system in which the spin can point in any direction in three dimensions, the order parameter is a vector 
\begin{equation}
\vec{m} = \frac{1}{N} \sum_i \vec{\sigma}_i,
\end{equation}
where $\vec{\sigma}_i = (\sigma_{ix}, \sigma_{iy},\sigma_{iz})$, for the standard Pauli spin matrices.

\begin{figure}
\begin{center}
\includegraphics[width=0.65\textwidth]{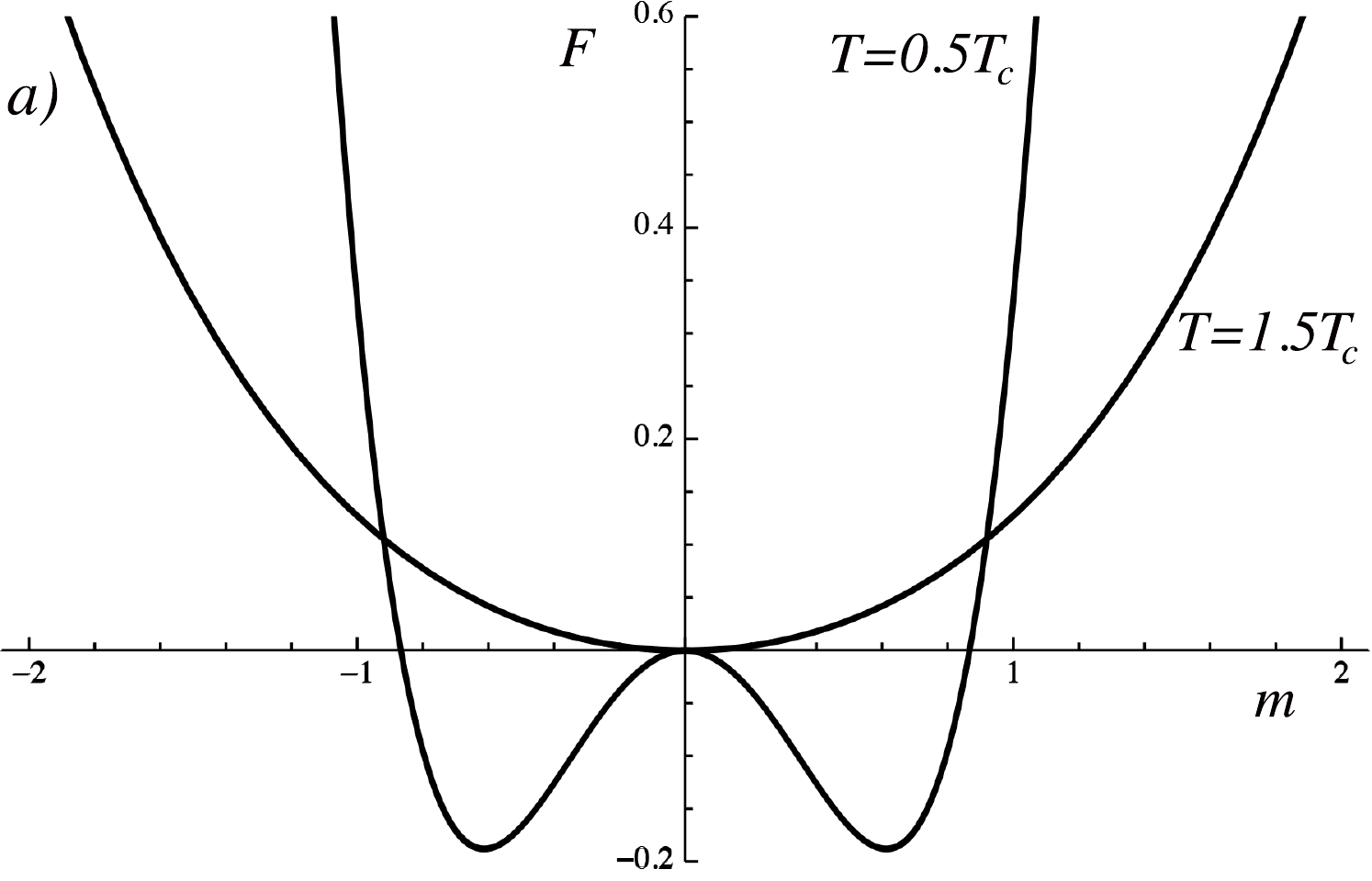}
\includegraphics[width=0.6\textwidth]{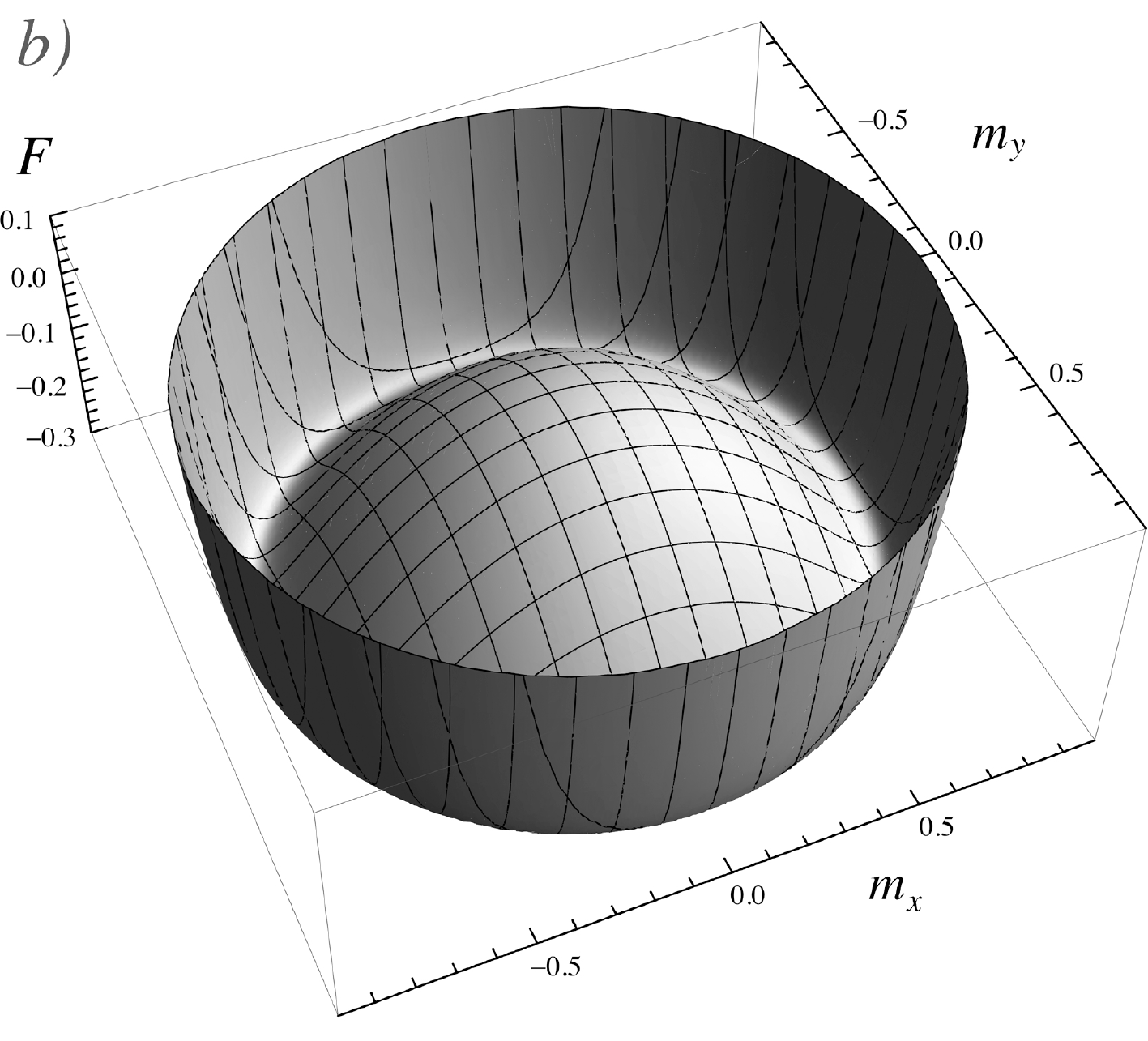}
\caption{a) Helmholtz free energy (\ref{ferrofree}) for the Ising model, for two temperatures, with $Nk_BT = 1$. b) Free energy profile for the case $T=0.5T_c$ with two degrees of freedom.}
\label{fig1}
\end{center}
\end{figure}

The mean-field solution for the free energy of (\ref{ising}) as a function of $m$ in the absence of external magnetic field can be exactly calculated \cite{snokech10}, and is 
\begin{eqnarray}
F &=& -Nk_BT \left(\ln2-\frac{T_c}{2T}m^2 + \ln \cosh \left(\frac{T_c}{T}m \right) \right) \nonumber\\
&\simeq& F_0+Nk_BT\left( \frac{T_c}{2T}\left(\frac{T-T_c}{T}\right)m^2 + \frac{1}{12}\left(\frac{T_c}{T}\right)^4 m^4  \right),
\label{ferrofree}
\end{eqnarray}
where $T_c$ is the critical temperature for the ferromagnetic phase transition. Figure~\ref{fig1}(a) shows this free energy for two temperatures, above and below $T_c$. As seen in this figure, at $T = T_c$ the shape of the curve switches from a single minimum at $m=0$ to two minima at finite $m$. The equivalent curve for a system which allows spin in two dimensions is the ``Mexican hat'' or ``wine bottle'' potential, illustrated in Figure~\ref{fig1}(b). The value of $m$ in equilibrium for a homogeneous system is found by solving 
\begin{eqnarray}
\frac{\partial F}{\partial m}  = Nk_BT\left( \frac{T_c}{T}\left(\frac{T-T_c}{T}\right)m + \frac{1}{3}\left(\frac{T_c}{T}\right)^4 m^3  \right) = 0.
\label{fmin}
\end{eqnarray}
The solution at $m=0$ is unstable when $T<T_c$.

The notion of spontaneous symmetry breaking can be seen by thinking of how the system behaves as the temperature passes through $T_c$ from above. The free energy is perfectly symmetric with respect to $m$, since there is no preferred direction for the spins in the absence of external magnetic field. Below $T_c$, the free energy curve remains symmetric, but the system can move to lower energy by breaking this symmetry, picking an energy minimum with $m \ne 0$.

How does the system choose a particular value of $m$ and not another? In condensed matter physics, it is quite easy to suppose that there is some stray magnetic field $B$ from outside the system which gives the system a kick in one direction or another. The system then amplifies this small asymmetry until it reaches a macroscopic average value of $m$. 

This type of spontaneous symmetry breaking is a model for numerous systems. For example, it can be applied to the onset of lasing. In this case, the control parameter is not the temperature, but the pump power, or optical gain. One writes the Maxwell's wave equation for the classical electric field $E$
\begin{equation}
-\omega^2E = \frac{\partial^2 E}{\partial t^2}+\frac{1}{\epsilon_0}\frac{\partial^2 P}{\partial t^2},
\label{max}
\end{equation}
where $P$ is the average polarization of the medium. For an ensemble of two-level quantum oscillators, one can write the polarization as
\begin{equation}
P(t) = {\sf Re}~ \tilde{d}\frac{N}{V} U_1,
\label{pol}
\end{equation}
where $\tilde{d}$ is an intrinsic dipole moment for the electric field coupling to the two-level oscillators, and $U_1$ is a component of the standard average Bloch vector $\vec{U} = (U_1,U_2,U_3)$, where
\begin{eqnarray}
U_1 &=& \langle a^\dagger_e a_g + a^\dagger_g a_e\rangle \nonumber\\
U_2 &=& i\langle a^\dagger_e a_g - a^\dagger_g a_e\rangle\nonumber\\
U_3 &=& \langle a^\dagger_e a_e - a^\dagger_e a_e \rangle,
\end{eqnarray}
for the excited ($e$) and ground ($g$) states of the two-level oscillator. Assuming the existence of a coherent electric field $E(t) = E_0e^{-i\omega t}$ and incoherent gain $G$, the Bloch equations for the evolution of this vector are
\begin{eqnarray}
\frac{\partial U_1}{\partial t} &=& -\frac{U_1}{T_2} +\omega_0U_2 -\omega_R U_3\sin\omega t \nonumber\\
\frac{\partial U_2}{\partial t} &=& -\frac{U_2}{T_2}-\omega_0U_1-\omega_RU_3 \cos\omega t \nonumber\\
\frac{\partial U_3}{\partial t} &=& -\frac{U_3+1}{T_1}+\omega_RU_1\sin\omega t +\omega_RU_2 \cos\omega t+ G(1-U_3),
\end{eqnarray}
where $T_1$ and $T_2$ are the relaxation and dephasing time constants, respectively, $\hbar \omega$ is the energy gap between the ground and excited states, and $\omega_R = \tilde{d}E/\hbar$ is the Rabi frequency, proportional to the electric field amplitude. 

Solving the Bloch equations in steady state for the amplitude of $U_1$ and using this in the polarization (\ref{pol}), which in turn is used in the Maxwell wave equation (\ref{max}), gives \cite{snokech11}
\begin{equation}
\frac{\partial E_0}{\partial t } = \frac{\omega}{2\epsilon_0}\left(AE_0 - BE_0^3\right),
\label{laser}
\end{equation}
where 
\begin{eqnarray}
A &=& \frac{\tilde{d}^2}{\hbar}\frac{N}{V}\left(\frac{G\tau-1}{G\tau+1}\right)\nonumber\\
B&=& A\tilde{d}^2\frac{\tau^2}{G\tau+1},
\end{eqnarray}
in which we have set $T_1 = T_2 = \tau$.

The steady-state solution of Equation (\ref{laser}) has the same form as (\ref{fmin}). When $A$ is positive, that is, when the gain exceeds a critical threshold, a small coherent part of the electric field will be amplified, growing in magnitude until it saturates at a nonzero value. If there is no coherent electric field, there will be nothing to amplify, but as in the case of the ferromagnet, we assume that some stray external field may impinge on the system. Since the $E_0 = 0$ point is unstable, like the $m=0$ point of the ferromagnet, any tiny fluctuation will cause it to evolve toward a stable point. 

This model of spontaneous symmetry breaking has also been applied to the early universe \cite{universe,kibble,zurek}. In this case, it is harder to imagine what might count as an ``external'' field that gives the tiny kick needed. Typically, in the condensed matter context, the nature of the tiny fluctuation is of little concern, since the instability of the symmetric point is assumed to always make it impossible for the system to remain there; at the unstable point the system will amplify even the tiniest fluctuation. With Bose condensates, however, the question has arisen whether there must be an external kick to bring about broken symmetry, and if so, where it comes from.

\section{Coherence in Condensates as Spontaneous Symmetry Breaking}

The above analysis can be mapped entirely to the case of Bose-Einstein condensation. The Hamiltonian in this case is
\begin{eqnarray}
H = \sum_{\vec{k}} E_k a^{\dagger}_{\vec{k}} a_{\vec{k}} + \frac{1}{2V}\sum_{\vec{p},\vec{q},\vec{k}} U(k) a^\dagger_{\vec{p}} a^\dagger_{\vec{q}}a_{\vec{q}+\vec{k}}a_{\vec{p}-\vec{k}}.
\label{BECH}
\end{eqnarray}
The Einstein \cite{einstein} argument ignores the interaction energy term and computes the total kinetic energy of a population of bosons. In this limit, one has simply
\begin{eqnarray}
N  =  \sum_{\vec{k}} \langle a^{\dagger}_{\vec{k}} a_{\vec{k}}\rangle = \sum_{\vec{k}} \langle N_{\vec{k}} \rangle 
\end{eqnarray}
where $N_{\vec{k}}$ is the occupation number of state $\vec{k}$. Converting the sum to an integral for a three-dimensional system, we have
\begin{equation}
N_{\rm ex} = \frac{V}{2\pi^2}\frac{m^{3/2}}{\hbar^3} \int_0^\infty \bar{N}(E) \sqrt{E} dE 
\label{EinsteinN}
\end{equation}
where $\bar{N} = 1/(e^{(E-\mu)/k_BT}-1)$ is the Bose-Einstein average occupation number. This gives the well known result that below a critical temperature $T_c$, the integral (\ref{EinsteinN}) cannot account for all the particles; there must be an additional population not accounted for by this sum over excited states, with zero kinetic energy, which we call the condensate.  

It has been pointed out by Nozieres \cite{noz} and others \cite{comb} that the Einstein argument does not address the stability of the condensate. 
In an infinite system, there is an infinite number of $k$-states near $k=0$ with negligible kinetic energy. In a non-interacting system, we could spread the condensate over any number of these states with negligible energy penalty. 

To see how the interactions affect the stability of the condensate, imagine that we have a system in which the kinetic interactions have already caused a macroscopic number of particles to accumulate in states with negligible kinetic energy near $\vec{k}=0$, but they are not in the same quantum state. We now imagine varying the fraction of this population of particles which is in the ground state, that is, the true condensate, and calculate the free energy of the system as a function of this fraction, using the Hamiltonian (\ref{BECH}). 

Computing this is nontrivial for a general interact potential $U(k)$, but we can see the general behavior if we assume $U(k) = U = $ constant. In this case, the expectation value of the Hamiltonian (\ref{BECH}) is 
\begin{eqnarray}
\langle H \rangle &=& 
 \frac{U}{2V}N_0^2 - \frac{U}{V}N_0\sum_{\vec{p}} N_{\vec{p}} +\frac{U}{2V}\sum_{\vec{p},\vec{q}} \left( \langle a^\dagger_{\vec{p}} a^\dagger_{\vec{q}}a_{\vec{q}}a_{\vec{p}}\rangle+\langle a^\dagger_{\vec{p}} a^\dagger_{\vec{q}}a_{\vec{p}}a_{\vec{q}}\rangle\right),\nonumber \\
\label{Hrestricted}
\end{eqnarray}
where $N_0$ is the number in the ground state. Neglecting terms in the last sum arising from commutation when $\vec{p} = \vec{q}$, since these will be small compared to the whole sum, this becomes
\begin{eqnarray}
\langle H \rangle &=& \frac{U}{2V}N_0^2 + \frac{U}{V}N_0N_{\rm ex} + \frac{U}{V}N_{\rm ex}^2 \nonumber\\
&=&  \frac{U}{V}\left( N^2 - NN_0 +\frac{1}{2}N_0^2\right),
\label{HN0}
\end{eqnarray}
where $N_{\rm ex} = \sum N_{\vec{k}} = N - N_0$, and $N$ is the total number of particles.
 
We can see already from (\ref{HN0}) that $\langle H \rangle$ will be lower when $N_0 > 0$.  To account for the entropy of particles leaving the condensate, we write the free energy
\begin{eqnarray}
F = \langle H \rangle - TS,
\end{eqnarray}
where 
\begin{eqnarray}
S &=& -k_B\sum_{\vec{k}} \left(N_{\vec{k}} \ln N_{\vec{k}} -(1+N_{\vec{k}})\ln(1+N_{\vec{k}}) \right) 
\end{eqnarray}
is the entropy of a boson gas \cite{bosonS}. We suppose that particles leaving the condensate move to a region in $k$-space near $k=0$ with $N_{\vec{k}} \gg 1$. Then we can approximate
\begin{eqnarray}
S &\simeq & k_B\sum_{\vec{k}} \frac{\partial }{\partial N_{\vec{k}}} N_{\vec{k}} \ln N_{\vec{k}} 
\simeq k_B\sum_{\vec{k}} \ln N_{\vec{k}}  .
\end{eqnarray}
Assuming that the value of $N_{\vec{k}}$ does not deviate too strongly from its average value $\bar{N}_{\vec{k}} = N_{\rm ex}/N_s$, where $N_s$ is the total number of states in the selected region of $k$-space, we obtain 
\begin{equation}
TS \simeq k_BT N_s \ln (N_{\rm ex}/N_s) = k_BTN_s [\ln (N-N_0)-\ln N_s].
\end{equation}
which allows us to write
\begin{eqnarray}
\frac{F}{V} &=& \frac{F_0}{V} - U\frac{N}{V}|\psi_0|^2 +\frac{1}{2}{U}|\psi_0|^4 -k_BT \frac{N_s}{V} [\ln (N/V-|\psi_0|^2)-\ln N_s/V],\nonumber\\
\label{BECfree}
\end{eqnarray}
where we have defined the wave function $\psi_0$ of the condensate as the order parameter, with $|\psi_0|^2 =N_0 /V $. 
We can estimate $N_s$, The number of states in the region around $k=0$ with $N_{\vec{k}} \gg 1$, using the relation
\begin{eqnarray}
N_s = \int_0^{E_{\rm cut}} D(E)dE ,
\end{eqnarray}
where $E_{\rm cut}$ is the energy at which $N_k \gg 1$, which we can take as $E_{\rm cut} \sim k_BT $. In three dimensions, this gives us 
\begin{eqnarray}
\frac{N_s}{V} &=& \frac{\sqrt{2}m^{3/2}}{2\pi^2\hbar^3}\int_0^{k_BT} dE \sqrt{E}  \nonumber \\
&=& \frac{\sqrt{2}(mk_BT)^{3/2}}{\pi^2\hbar^3} \sim \frac{1}{\lambda_{dB}^3} ,
\end{eqnarray}
where $\lambda_{dB}$ is the deBroglie wavelength determined by setting $(\hbar k)^2/2m = k_BT$, with $k = 2\pi/\lambda_{dB}$.

\begin{figure}
\begin{center}
\includegraphics[width=0.7\textwidth]{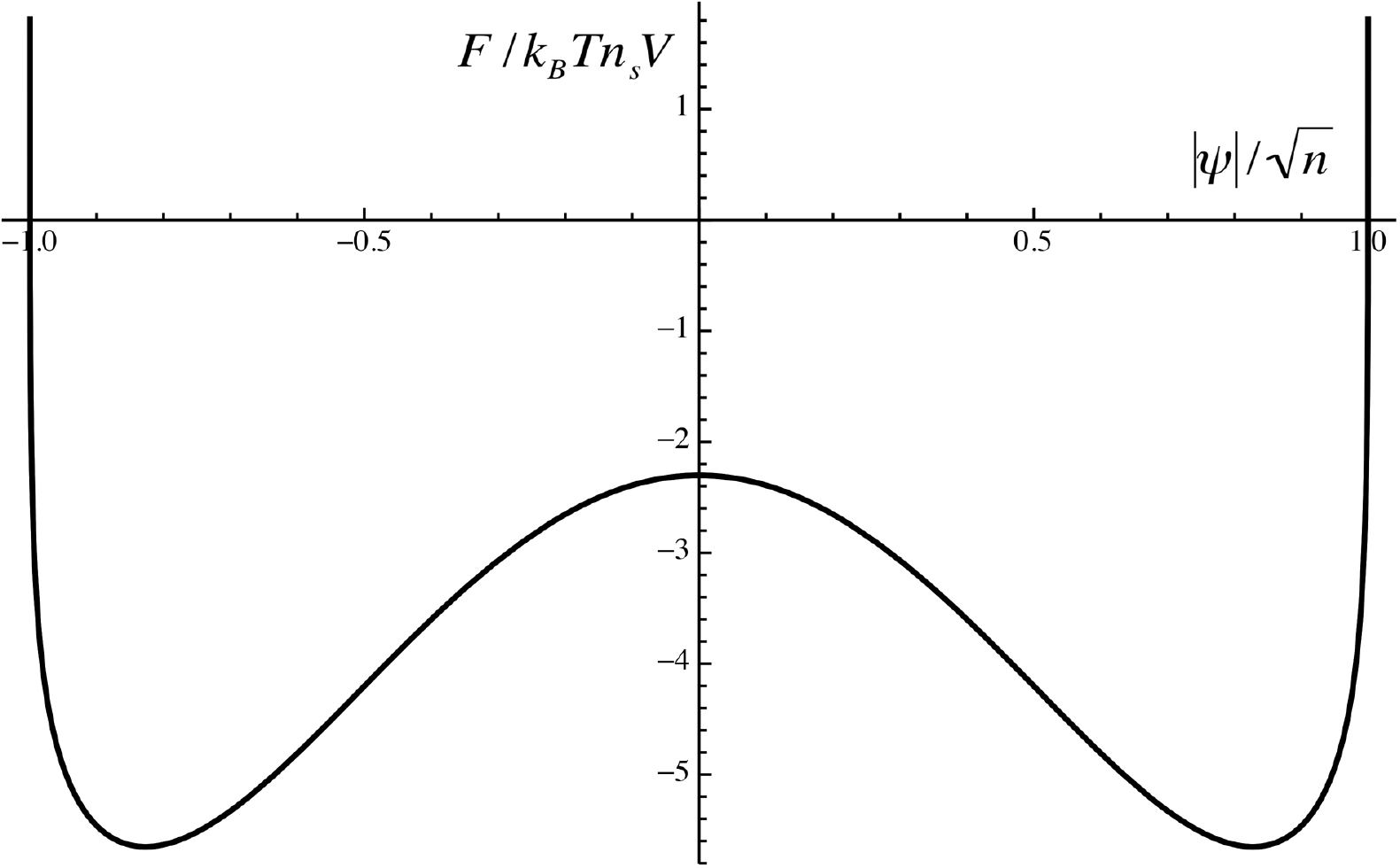}
\caption{a) Free energy for an interacting Bose-Einstein condensate using the approximation (\ref{BECfree}), with $Un/k_BT = 1$ and $n/n_s = 10$, where $n = N/V$ is the particle density and $n_s = N_s/V$.}
\label{fig2}
\end{center}
\end{figure}

The free energy (\ref{BECfree}) is plotted in Figure~\ref{fig2}, which has the same generic form as Figure~\ref{fig1}(a), and will also have the two-dimensional form of Figure~\ref{fig1}(b), but here the two components are the real and imaginary parts of $\psi_0$. The symmetric center point at $\psi_0 =0$ is unstable, because increasing the number of particles in the condensate reduces the interaction energy. This occurs because exchange in a bosonic system favors having particles in the same state. This is true for composite bosons as well \cite{comb}. Eventually the entropy cost of adding particles to the zero-entropy condensate state will prevent all the particles from entering it. The stable value is obtained on a ring with fixed amplitude, that is, $\psi_0 = \sqrt{N_0}e^{i\theta}$, where $\theta$ is arbitrary. Zero-energy variation of $\theta$ is known as a Goldstone mode, whereas oscillation of $N_0$ in the radial direction is known as a Higgs mode.

The depletion of the condensate depends on the ratio of the interaction strength $Un$ to $k_BT$ and the average occupation number of the non-condensed region. The stable point will move closer to $N_0 = N$ for stronger interactions and higher degeneracy of the excited particles. Approximating $\ln (N-N_0)$ at the stable point $N_0 = N_{eq}$ as 
\begin{equation}
\ln (N-N_0) \simeq \ln (N-N_{eq}) - \frac{N_0-N_{eq}}{N-N_{eq}},
\end{equation}
and taking the first derivative of $F$ with respect to $\psi^*_0$, we obtain
\begin{equation}
-\left(\frac{U}{V}N + \frac{k_BT N_s}{N-N_{eq}}\right)\psi_0 +\frac{U}{V}|\psi_0|^2\psi_0 = 0.
\end{equation}
This has the same form as  
(\ref{laser}), and is the same as the Landau-Ginzburg equation for a homogeneous boson gas.   

When we look at nonequilibrium behavior, the symmetry breaking in a Bose-Einstein condensate also has similarities with the laser system. Starting with the same Hamilitonian (\ref{BECH}) and deriving the quantum Boltzmann equation using second-order perturbation theory, the equation for the evolution of a coherent state in the ground state of the bosons is found to be \cite{girv1,girv2}
 \begin{eqnarray}
&&\frac{d}{dt} \langle a_0 \rangle = \langle a_0 \rangle \frac{\pi }{\hbar}\left(\frac{2U}{V}\right)^2
\sum_{\vec{p},\vec{q}}
 \left[
\langle \hat{N}_{\vec{q}}\rangle \langle \hat{N}_{\vec{p}-\vec{q}}\rangle (1+ \langle \hat{N}_{\vec{p}}\rangle)
\right.  \nonumber\\
&&\hspace{1.5cm}\left.
-\langle \hat{N}_{\vec{p}}\rangle(1+ \langle \hat{N}_{\vec{q}}\rangle)(1+\langle \hat{N}_{\vec{p}-\vec{q}}\rangle) \right]   \delta(E_{\vec{p}}-E_{\vec{q}}-E_{\vec{p}-\vec{q}}) .\nonumber\\
\label{BECphase}
 \end{eqnarray}
The first term in the square brackets gives the total in-scattering rate, and the second term is the total out-scattering rate.  

As with the electric field in the case of a laser controlled by Equation (\ref{fmin}), we imagine a tiny coherent part has already been created somehow, and then see that the system can amplify this coherent part until it becomes macroscopic. Equation~(\ref{BECphase}) implies that this amplification will occur whenever
there is net influx into the ground state, which occurs when the system approaches the BEC equilibrium kinetic-energy distribution \cite{snokewolfe,boltz}. This implies exponential growth of the amplitude of a phase-coherent part; this growth will end when the system reaches equilibrium, and the influx to the ground state and the outflow balance.

Because the equations for growth of the condensate have the same form as those commonly used for spontaneous symmetry breaking in other systems, it is natural to assume that the same thing occurs in the case of Bose-Einstein condensation. However, many in the field of condensates have paid attention to the crucial role of the fluctuation which seeds the condensate. In the case of a ferromagnet or laser, it is easy to imagine that there is a stray magnetic field or electric field from outside the system. In the case of a condensate, however, the amplitude $a_0$ (or $\psi_0$ in the spatial domain) corresponds to the creation or destruction of particles, and in a strictly number-conserved system there is no external field that couples into the system to do this. The system is analogous to the early-universe scenario of spontaneous symmetry breaking, in which there appears to be no ``outside'' to give the small kick to break the symmetry.

On the other hand, we do not strictly know that there is no coupling term to the order parameter in a matter wave such as a cold atom condensate. If the proton decayed every $10^{30}$ years, we would have number conservation for all intents and purposes, but there would still be a tiny term that could give a fluctuation which is amplified. 

\section{Coherence in Condensates as a Measurement Phenomenon}
\label{sect.meas}

As a solution to this problem, the predominant approach in the cold atom community is to view spontaneous symmetry breaking as one of many possible descriptions for the system. The question is most commonly posed in the case of interference experiments, where two Bose-Einstein Condensates are released and the resulting interference patterns are measured \cite{Andrews1997}. An interference pattern between two condensates seems to imply a definite phase relation between the two, which could be most straight-forwardly described with each condensate in a coherent state. The problem arises, however, that the standard definition of a coherent state \cite{glauber} has an indefinite number of particles. At $T=0$, however, a condensate has a definite number of particles, namely all the particles in the system. Does this then imply that at $T=0$, in a truly isolated system, there would be no interference pattern? Of course, all real experiments are performed at finite temperature, so that there is fluctuation in the number of condensate particles. Do we expect that in the $T\rightarrow 0$ limit there will always be some tiny non-number-conserving term which gives the needed kick into a coherent state?

In the experiments performed, it was unknown whether the condensates had a well-defined phase before the measurement was made, or only afterward. In the spontaneous symmetry breaking scenario, one can suppose that each condensate spontaneously could acquire a well-defined amplitude prior to interacting. But the experiments which have been performed allow for the possibility that the amplitude was well-defined only after the measurement was made. 

There are two related setups in which this has been discussed at length: one is where two condensates that are are later measured are coupled with each other, e.g., in the case of Bose-Einstein condensates in a double-well with fixed total particle number \cite{Wong1996,Wright1996,Imamoglu1997,Javanainen1997,Milburn1997}, and the other involves the interference of two independent condensates, each initially with a fixed total particle number \cite{Naraschewski1996,Castin1997,Javanainen1996,Cirac1996}. In each case, it can be shown that number-conserving approaches give rise to the same experimental predictions for interference patters as the assumption of spontaneous symmetry breaking. This can be extended beyond interference experiments to a range of processes described by a Gross-Pittaevskii equation or Bolgoliubov theory in a number conserving approach \cite{Gardiner1997,Castin1998,Leggett2001}. 

In the case of the double well, the well-defined phase description could be applied with a justification that the exchange of particles between the coupled condensates leads to an indeterminate number between the individual wells. In both cases, however, it can be equally assumed that the phase of the sample is indeterminate until the condensate is actually measured. If two condensates with fixed particle number are released, and the resulting interference pattern is detected by absorption imaging, then the first particle that is measured could have come from either of the two independent condensates. This measurement process sets up a superposition state between the two condensates, and it is shown in Refs.~\cite{Naraschewski1996,Castin1997,Javanainen1996,Cirac1996} that a proper analysis of the resulting continuous measurement process gives rise to the same interference pattern that would be expected from symmetry-broken condensates with well-defined (but unknown) initial phases. This same argument, that number states can give rise to interference identical to that in a fixed-phase representation (e.g., a coherent state) was applied to optical coherence by M{\o}lmer \cite{Molmer1997,Molmer1997b}. He summarized these ideas in a short poem for the abstract of Ref.~\cite{Molmer1997}: ``Coherent states may be of use, so they say, but they wouldn't be missed if they didn't exist.'' The standard approach in optics, however, is to view coherent states as physically real. This view is supported by the fact that in systems without number conservation, pure number states (a.k.a.~Fock states) are unstable to becoming coherent states \cite{girv2}, by the same type of calculation which led to Equation (\ref{BECphase}) above.  Contra M{\o}lmer, the standard approach treats {\em number} states as mostly dispensable, as for example in this quote from a standard laser textbook \cite{siegman}: ``We have hardly mentioned photons yet in this book...The problem with the simple photon desciption... is that it leaves out and even hides the important wave aspects of the laser interaction process.'' Quantum mechanics, of course, allows either coherent states or Fock states as bases.

A simple example for comparison between the symmetry-broken description with coherent states and a description that conserves the number of particles is the case of collapse and revival of matter-wave interference in an optical lattice. For atoms confined in 3D, this was first realised by Greiner et al. in 2002 \cite{Greiner2002}, and then studied in more detail by Will et al.~\cite{Will2010}. The basic context is that a very weakly interacting Bose-Einstein condensate is loaded into a 3D optical lattice. We can then see how this depends on atom number and system size, taking the lattice to have $M$ sites, and $N$ homogeneously distributed atoms. These sites are initially coupled by tunnelling, but this is switched off by suddenly making the lattice deep. On-site energy shifts that are dependent on the particle number then dephase correlations between different sites, leading to a collapse in interference peaks when the atoms are released from the lattice.

In the spontaneous symmetry breaking case, the picture is very clear. We can write the initial state of the non-interacting BEC as a product of coherent states in the local particle number,  
\begin{equation}
|\mathrm{BEC}_{\mathrm{SB}}\rangle= \prod_l^M e^{-\frac{|\beta_l|^2}{2}}
e^{\beta_i b_l^\dag} |\mathrm{vac}\rangle,
\label{eq:naive_ansatz}
\end{equation}
where $\beta_l$ is the mean particle number on site $l$.

On-site energy shifts due to the number of particles, can be described by an interaction Hamiltonian $H_{\rm int}=(U/2)\sum_l \hat n_l (\hat n_l -1)$, where $\hat n_l=b_l^\dag b_l$ is the number operator for bosons on site $l$, $b_l$ is the annihilation operator for a bosonic atom on site $l$, and $U$ is the two-particle collisional energy shift. As a function of time, we can then see that the correlations behave as
\begin{eqnarray}
\langle b_{i}^{\dag} b_{j} \rangle_t = \langle b_i \rangle_t^* \langle
b_j\rangle_t  \notag = \beta_i^* \beta_j e^{ |\beta_i|^{2}(e^{iUt}-1) } e^{
|\beta_j|^{2}(e^{-iUt}-1) } = |\beta|^2 e^{ |\beta|^{2}(2\cos(Ut)-2) },\nonumber \\
\label{eq:coherent_spdm}
\end{eqnarray}
with $\beta_l=\beta$ in the homogeneous system. 

In the alternative case, with fixed particle number, we van write the initial state of $N$ particles in a homogeneous system of $M$ sites as
\begin{align}  
|\mathrm{BEC}_N\rangle = \frac{1}{\sqrt{N! N^N}} \left( \sum_i^M \beta
\hat b_i^\dag \right)^N |\mathrm{vac}\rangle.
\end{align}
It is helpful to rewrite this state as a number projection of the coherent states. We can see how this is possible by writing 
\begin{align}  \label{eq:betai_state}
|\{ \beta \} \rangle &\equiv \prod_{k}^{M} | \beta \rangle_k \equiv
\prod_{k}^{M} e^{ -\frac{|\beta|^{2}}{2} } e^{ \beta b_{k}^{\dag} } |%
\mathrm{vac}\rangle,
\end{align}
where with a projection operator $P_N$, we obtain 
\begin{align}
P_N |\{ \beta \} \rangle &= e^{ -\frac{1}{2} \sum_k^M |\beta|^2 }
P_{N} e^{ \sum_{k}^{M}\beta b_{k}^{\dag} }|\mathrm{vac}\rangle  \notag \\
& = e^{ -\frac{1}{2} \sum_k^M |\beta|^2 } P_{N} \sum_{n=0}^\infty \frac{1%
}{n!} \left( \sum_{k}^M \beta b_k^\dag \right)^n |\mathrm{vac}\rangle 
\notag \\
&= \frac{1}{\sqrt{\mathcal{N}}} e^{ -\frac{1}{2} \sum_k^M |\beta|^2 } 
\frac{1}{N!} \left( \sum_{k}^M \beta b_k^\dag \right)^N |\mathrm{vac}%
\rangle,
\end{align}
which is identical to the ground state with fixed particle numbers. We now make use of a trick in which we write the Fock state of fixed particle number as a phase averaged coherent state \cite{Javanainen1996},
\begin{align}  \label{eq:projection_operator}
P_{N} &= \frac{1}{2\pi \sqrt{\mathcal{N}} } \int_{-\pi}^{\pi} d\phi\, e^{ i(%
\hat{N}-N)\phi } ,
\end{align}
where the total particle number operator is given by $\hat{N} = \sum_{i} b_{i}^{\dagger}
b_{i}$. 
Writing the state with fixed total particle number in this form means that the time-evolved SPDM can then be written exactly as 
\begin{align}
F \langle b_{i}^{\dag} b_{j}\rangle_t &= \langle\mathrm{BEC}_{N}|e^{iHt}
b_{i}^{\dag} b_{j} e^{-iHt}|\mathrm{BEC}_{N}\rangle  \notag \\
&= \frac{1}{2\pi \mathcal{N} } \int_{-\pi}^{+\pi} d\phi\, e^{-iN\phi} \langle \{ \beta \} | e^{iHt} b_{i}^{\dag} b_{j} e^{-iHt}
e^{\hat N \phi} | \{ \beta \} \rangle  \notag \\
& = \frac{1}{2\pi \mathcal{N}} \int_{-\pi}^{+\pi} d\phi\, e^{-i N\phi}  \langle \{ \beta \} | e^{iHt} b_{i}^{\dag }b_{j} e^{-iHt}
| \{ \beta e^{i\phi } \} \rangle .  \label{eq:spdm_integral}
\end{align}


In this integral, the diagonal elements of the SPDM remain
constant in time, where, as expected from the density, $\langle b_{i}^{\dag} b_{i}\rangle_{t}= |\beta|^2$. The key to understanding the time dependence of the interference pattern is to evaulate the off-diagonal elements, which factorise as
\begin{align}
& \langle \{ \beta \} | e^{iHt} b_{i}^{\dag} b_{j} e^{-iHt} | \{ \beta
e^{i\phi} \} \rangle  \notag \\
=& \quad  \langle \beta | e^{iU\hat n_i (\hat n_i-1)t/2} b_{i}^{\dag} e^{-iU\hat n_i (\hat n_i-1)t/2} | \beta
e^{i\phi} \rangle_{i}\notag \\
& \quad \times  \langle \beta | e^{iU\hat n_i (\hat n_j-1) t/2} b_{j}e^{-iU\hat n_i (\hat n_j-1) t/2} |
\beta e^{i\phi}\rangle _{j}  \prod_{k\neq i,j} \langle \beta | \beta e^{i\phi}
\rangle_k .
\end{align}
We can then compute  
\begin{align}
&\langle \beta | e^{iU\hat n_i (\hat n_j-1) t/2} b_{j}e^{-iU\hat n_i (\hat n_j-1) t/2} | \beta
e^{i\phi}\rangle _{j} = \beta e^{i\phi} e^{-i\varepsilon_j t} e^{
|\beta|^{2} \left[ e^{i\phi} e^{-iUt}-1 \right] } ,
\label{eq:factorized_elements}
\end{align}
and 
\begin{align}
& \langle \beta | e^{iU\hat n_i (\hat n_i-1)t/2} b_{i}^\dag e^{-iU\hat n_i (\hat n_i-1)t/2} |\beta
e^{i\phi}\rangle_{i} = \beta^* e^{i\varepsilon_i t} e^{ |\beta|^{2} 
\left[ e^{i\phi} e^{iUt}-1 \right] } , \label{eq:factorized_elements2}
\end{align}
so that for $i\neq j$,
\begin{align}\label{eq:spdm_evolution_full}
\langle b_{i}^{\dag} b_{j}\rangle_t &= \frac{\beta^* \beta}{2\pi 
\mathcal{N}} \, e^{-N} e^{i (\varepsilon_i - \varepsilon_j) t}
\int_{-\pi}^{+\pi}d\phi \, e^{-i(N-1)\phi}  \\
&\quad \times e^{ e^{i\phi} \left [ |\beta|^2 e^{-iUt} + |\beta|^2
e^{iUt} + \sum_{k \neq i,j} |\beta |^2 \right] }  \notag \\
&= | \beta|^2 \left [ 1 + \frac{|\beta|^2%
}{N} \left( 2\cos{(Ut)} - 2 \right) \right]^{(N-1)} . \label{eq:spdm}
\end{align}

In the limit where the particle number is much
larger than the onsite density, i.e., $N \gg |\beta|^2$, we find 
\begin{align}
|\beta|^2 \left[ 1 + \frac{|\beta|^2}{N} \left( 2 \cos(Ut) - 2 \right) %
\right]^{(N-1)} \overset{N \gg |\beta|^2}{\longrightarrow} |\beta|^2 e^{ |\beta|^2 \left[ 2
\cos(Ut) - 2 \right] } ,
\end{align}
which reproduces the result from coherent states in Eq.~(\ref{eq:coherent_spdm}).

\begin{figure}[h]
\begin{center}
\includegraphics[width=0.7\textwidth]{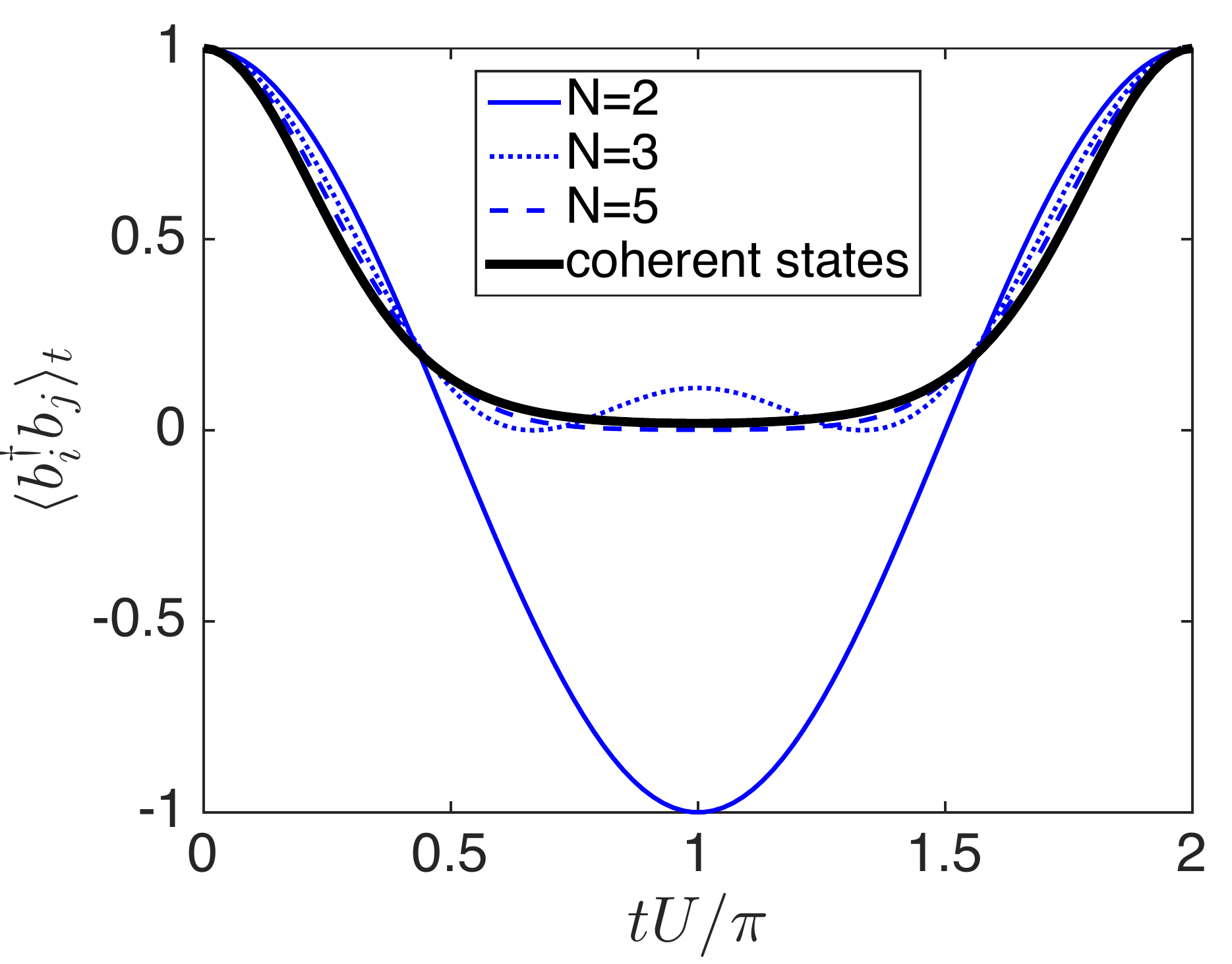}
\end{center}
\caption{Plot of the time-evolution of the off-diagonal elements of the single-particle density matrix
in a homogeneous system (i.e., periodic boundary conditions) at unit average filling. We clearly see the first collapse and revival of these correlation functions, and compare the predictions from the symmetry-broken solution of  Eq.~(%
\protect\ref{eq:coherent_spdm}) (solid heavy line) and the number-conserving calculation in Eq.~(%
\protect\ref{eq:spdm}). Redrawn from 
Ref.~\cite{Schachenmayer2011}.}
\label{fig:spdm_evolution_n}
\end{figure}

To see how the interference pattern observed in an experiment depends on the system size and particle number, we compute the height of the zero-quasimomentum peak, $n_{q=0} = (1/M ) \sum_{i,j}^M
\langle b_{i}^{\dag} b_{j}\rangle_t = (M - 1) \langle b_{i}^{\dag} b_{j}
\rangle + |\beta|^2$. This corresponds to the visibility of an interference pattern after a long time of flight. In Fig.~\ref{fig:spdm_evolution_n} we plot  show a plot of the time-evolution of the these visibilities beginning from a ground state with density $|\beta|^2=1$. As the particle number increases, we see that the values converge rapidly to the values from coherent states: already for $N \sim 5$, the results are difficult to distinguish from each other. Defining the relative difference, we can show that the difference decreases proportional to $1/M$ \cite{Schachenmayer2011}.

As a final comment on measurement-based treatments with fixed initial numbers, we note that one way to look at the standard spontaneous symmetry breaking scenario is to say that some form of outside kick has perturbed the system. In the case discussed here, or also in interference of independent condensates, one might suggest that the outside influence is the interaction with the measuring apparatus. This raises the question, addressed by a whole host of philosophy, of what measurement really is. Perhaps the measurement itself involves spontaneous symmetry breaking of the standard kind, when a stray field external to the detector causes it to respond to a matter wave with a collapse in on direction or another. But many studies have shown that if this is the case, some degree of nonlocality must also enter. Detectors seem to coordinate their responses across light-like separations.  

\section{Spontaneous Symmetry Breaking in Photon and Polariton Condensates}

As we have seen in Section~\ref{sect.elem}, the standard model of spontaneous symmetry breaking envisions a small fluctuation which is amplified. In the case of photon \cite{weitz} and polariton condensates \cite{deveaud,snoke1,snoke2,baum,bloch-JJ}, this fluctuation can come from an external electromagnetic field. 

A polariton condensate is fundamentally no different from a photon condensate; we can see this in the following derivation of the polariton wave equation.  We start with Maxwell's wave equation in a nonlinear isotropic medium,
\begin{equation}
\nabla^2E = \frac{n^2}{c^2}\frac{\partial^2 E}{\partial t^2} + 4\mu_0\chi^{(3)}\frac{\partial^2 }{\partial t^2}|E|^2E,
\label{maxwell}
\end{equation}
where $\chi^{(3)}$ is the standard nonlinear optical constant, and we ignore frequency-mixing terms in the general $E^3$ nonlinear response. In the standard polariton scenario, this nonlinear term is produced by a sharp electronic resonance, namely a two-level oscillation. In condensed matter systems, this is usually an excitonic excitation of the valence band electrons.

Writing the solution in the form $
E =\psi(x,t) e^{-i\omega t}$, 
and keeping only leading terms in frequency (known as the slowly varying envelope approximation), we have for the time derivative of $E$,
\begin{eqnarray}
\frac{\partial^2 E}{\partial t^2} 
&\simeq&  \left(-\omega^2\psi -2i\omega\frac{\partial \psi}{\partial t}\right)e^{-i\omega t},  
\end{eqnarray}
and for the time derivative of the nonlinear term
\begin{eqnarray}
\frac{\partial^2 }{\partial t^2} |E|^2E 
&\simeq& 
-\omega^2|\psi|^2\psi e^{-i\omega t}.  
\end{eqnarray}

The standard polariton structure uses a planar or nearly-planar cavity to give one confined direction of the optical mode. We therefore distinguish between the component of momentum $k_z$ in the direction of the cavity confinement, which is fixed by the cavity length, and the momentum $k_{\|}$ for motion in the two-dimensional plane perpendicular to this direction, which is free. We therefore write $
\psi = \psi(\vec{x})e^{i(k_{\|}\cdot\vec{x}+k_zz)}.
$
The full Maxwell wave equation (\ref{maxwell}) then becomes
\begin{eqnarray}
&& (-(k_z^2+k_{\|}^2)\psi + \nabla_{\|}^2\psi)\nonumber\\
&&\hspace{1cm}
=(n/c)^2\left(-\omega^2\psi -2i\omega\frac{\partial \psi}{\partial t}\right)-4\mu_0\chi^{(3)}
\omega^2|\psi|^2\psi  
. 
\end{eqnarray}
Since $\omega^2 = (c/n)^2(k_z^2 + k_{\|}^2)$, this becomes
\begin{eqnarray}
&& \nabla_{\|}^2\psi 
=(n/c)^2\left(-2i\omega\frac{\partial \psi}{\partial t}\right)-4\mu_0\chi^{(3)}
\omega^2|\psi|^2\psi  .
\label{opticalGP}
\end{eqnarray}

Near $k_{\|} = 0$, we can approximate
\begin{equation}
\hbar\omega  =\hbar(c/n) \sqrt{k_z^2 + k_{\|}^2} \ \simeq  \ \hbar(c/n)k_z \left(1+\frac{k_{\|}^2}{2k_z^2}\right)
\equiv \hbar\omega_0 + \frac{\hbar^2k_{\|}^2}{2m},
\end{equation}
which gives an effective mass for the photon motion in the plane. 
For the first term on the right-hand side of (\ref{opticalGP}), we approximate
\begin{equation}
\omega \simeq \omega_0 = \frac{m(c/n)^2}{\hbar}.
\end{equation}
Therefore we can rewrite (\ref{opticalGP}) as
\begin{equation}
i\hbar\frac{\partial \psi}{\partial t} = 
-\frac{\hbar^2}{2m} \nabla_{\|}^2\psi - \frac{2\mu_0\chi^{(3)}(\hbar\omega)^2}{m}|\psi|^2\psi,
\label{GP1}
\end{equation}
which we can rewrite as
\begin{equation}
i\hbar\frac{\partial \psi}{\partial t} = 
-\frac{\hbar^2}{2m} \nabla_{\|}^2\psi + U|\psi|^2\psi.
\label{GP2}
\end{equation}
This is a Gross-Pitaevskii equation, or nonlinear Schr\"odinger equation. Note that although the Maxwell wave equation is second order in the time derivative, this equation is first order in the time derivative, as in a typical Schr\"odinger equation.

The polariton and photon condensates therefore can follow the standard scenario of spontaneous symmetry breaking as occurs in a laser, in which one imagines a small stray electromagnetic field which is amplified. The equations which govern the polariton condensate, however, are identical to those of a standard condensate. In general in polariton condensates one can add generation and decay terms to the Gross-Pitaevskii equation (\ref{GP2}), as done by Carusotto and Keeling and coworkers \cite{caru,keeling} but this distinction from standard condensates has become less significant in recent years.  On one hand, to be strictly accurate, the same type of term should be written for cold atom condensates, because these condensates have particle loss mechanisms due to evaporation from their traps. On the other hand, the lifetime of polaritons in microcavities has been steadily increasing, so that the ratio of the lifetime of the particles in the system to their collision time can be several hundred, comparable to the ratio for cold atoms in traps. Therefore in both the cold atom and polariton condensate systems, it is reasonable to drop the generation and decay terms as negligible in many cases.

The fact that polaritons decay into photons which leak out of the cavity mirrors means that it is possible to directly observe the phase amplitude of the polaritons in interference measurements. In this case, the interference is not between two condensates, but is between two different regions of the same condensate, more similar to the interference in multi-well systems discussed above. 
These results are discussed further in Chapter ....

\section{Conclusions}

A question one can ask is whether condensates such as polariton and photon condensates (and magnon condensates \cite{magnon}) can be viewed as ``real'' condensates, if they are known to have weak coupling to the outside world which allows spontaneous symmetry breaking of the ferromagnetic type. On this, it would seem strange to treat them as entirely different phenomena when the equations governing their behavior, once the symmetry has been broken, are identical to those governing the behavior of atom condensates. 

Another question one can ask is whether {\em all} spontaneous symmetry breaking must intrinsically be of the ferromagnetic type; that is, whether there must be some external fluctuation, no matter how tiny, which is amplified by the instability of the system. On this, it is clear that a scenario in which there is no outside field to break the symmetry can still result in broken symmetry when quantum measurement is taken into account. However, since we do not fully understand the measurement process, this may simply beg the question, because it cannot be ruled out that measurement itself involves nonlocal broken symmetry somehow in the measuring apparatus. 

All of this thinking applies to the cosmology of the early universe. While it is clear that scenarios exist in which a state with broken symmetry can have lower energy than a symmetric state, we do not know how to introduce an ``external'' fluctuation to break the symmetry in the ferromagnetic analogy, but it is also hard to apply the measurement-broken-symmetry scenario to the early universe, without knowing what might count as an observer.

\bibliography{ubec_daley}
\bibliographystyle{cambridgeauthordate}

\end{document}